\documentclass[%
 reprint,
 amsmath,amssymb,
 aps,
 pre,
 floatfix,   % ← uncomment this
]{revtex4-2}

\usepackage{graphicx}% Include figure files
\graphicspath{{Figures/}{Templates/}}
\DeclareGraphicsExtensions{.pdf,.png,.jpg,.jpeg}
\usepackage{dcolumn}% Align table columns on decimal point
\usepackage{bm}% bold math
\usepackage{soul}
\usepackage{color}
\usepackage{makecell}
\usepackage{url}

\begin{document}

\preprint{APS/123-QED}

\title{Physically-Motivated Primitive Path Analysis of Entangled Polymer Networks}

% Force line breaks with \\

\author{B M Shahi Sifat Mottaqin}
\author{Benjamin Morrow}%
\author{Robert J. Wagner}%
 \email{Robert.J.Wagner@Binghamton.edu}

\affiliation{%
 Department of Mechanical Engineering\\Thomas J. Watson College of Engineering and Applied Science\\Binghamton University, Binghamton, NY, USA
 % \textbackslash\textbackslash
}%


%\affiliation{
%\affiliation{
%\affiliation{%


\date{\today}% It is always \today, today,

%%%%%%%%%%%%%%%%%%%%%%%%%%%%%%%%%%%%%%%%%%%%%%%%%%%%%%%%%%%%%%%%%%%%%%%%%%%%%%%%%%%%%%%%%%%%%%%%%%%%%%%
\begin{abstract}
Physical entanglements between polymer chains demonstrably enhance the modulus, strength, and toughness of elastomers and gels. However, it remains difficult to directly relate entanglement micromechanics to their macroscopic mechanical benefits. Experimentally investigating entanglements is difficult due to their nanoscale sizes, subsurface locations, and chemical indistinguishability from their surroundings. Computationally mapping structure-property relations remains difficult due to the high cost of predicting macroscale properties using physics-based models that enable direct entanglement observation, such as coarse-grained molecular dynamics (CGMD). Furthermore, entanglements are transient, configuration-dependent features without clear quantitative definitions. To address this ambiguity, we introduce an approach that quantitatively defines local entanglements along the backbones of simulated polymers based on the Gaussian Linking Number. We further introduce a geometric center of entanglement and verify that it represents the position through which entropic chain forces are transmitted using Kremer-Grest CGMD simulations. Unlike existing approaches, which output a single linking number for pairs of chains, our approach identifies the multitude of load-transmitting, inter- and intra-chain entanglements along a polymer's backbone. Towards bridging scales, we introduce a topological distillation algorithm that converts entangled CGMD networks into representative discrete network models (DNMs), representing entanglements as vertices and primitive paths between them as load-transmitting edges. Our DNMs reproduce small-strain virial stress predictions of the Kremer-Grest model with a 97\% reduction in computational cost, verifying the physical accuracy and cost benefits of this approach. Moving forward, this distillation procedure will facilitate physics-based, predictive modeling of entangled network mechanics, from polymers to architected metamaterials.
\end{abstract}

\maketitle

%\tableofcontents

%%%%%%%%%%%%%%%%%%%%%%%%%%%%%%%%%%%%%%%%%%%%%%%%%%%%%%%%%%%%%%%%%%%%%%%%%%%%%%%%%%%%%%%%%%%%%%%%%%%%
\section{\label{sec:Introduction}INTRODUCTION}

\begin{figure}[!htbp]
        \includegraphics[width=\columnwidth]{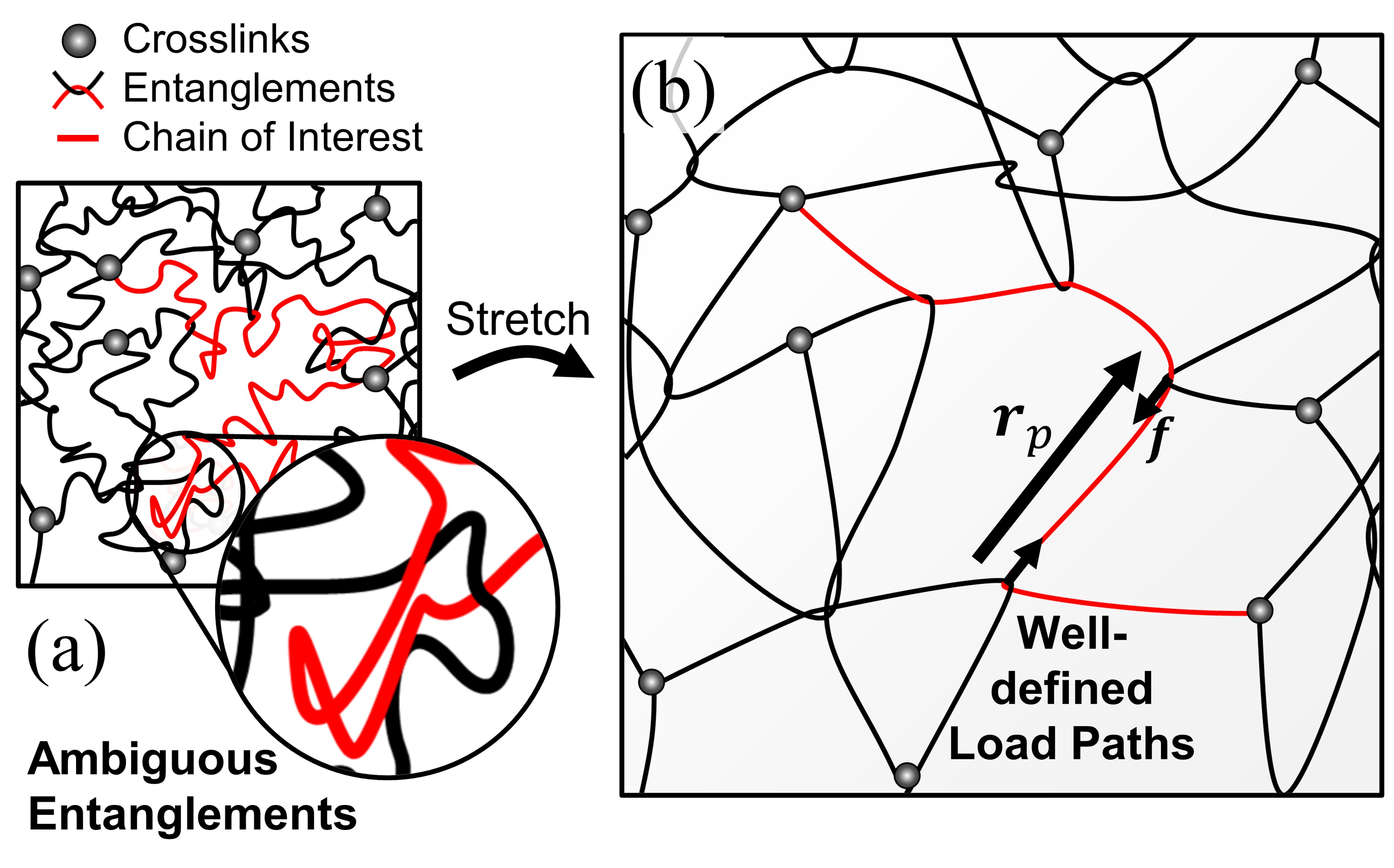}
        \caption{\label{fig:Primitive_Path_Figure}
        \textbf{Primitive Paths.} \textbf{(a-b)} Illustrations of a 2D, crosslinked polymer network at \textbf{(a)} low and \textbf{(b)} high biaxial stretch. \textbf{(a)} Close up of a highly coiled entanglement illustrates the topological ambiguity of entanglements at low stretch. \textbf{(b)} A primitive path vector, $\bm r_p$, between two entanglements is denoted along with the corresponding chain force, $\bm f$, acting at the entanglements. Load paths are relatively intuitive to identify in highly stretched networks.}
    \end{figure}

The mechanical properties of polymer melts, elastomers, and gels are predominantly governed by the entropic elasticity of their underlying chains and the load-transmitting interactions between them \cite{flory_principles_1953,rubinstein_polymer_2003}. Polymer chains in elastomeric networks primarily transmit load to one another through crosslinks or -- at sufficiently high molecular weights -- physical entanglements \cite{rubinstein_polymer_2003}. While crosslinks are generally well-defined junctions that transmit inter-chain loads wherever two chains are chemically or physically bonded, entanglements are ambiguous structures defined by transient coiling of neighboring chains (Fig.~\ref{fig:Primitive_Path_Figure}(a)). 

    \begin{figure*}[t]
        \includegraphics[width=\textwidth]{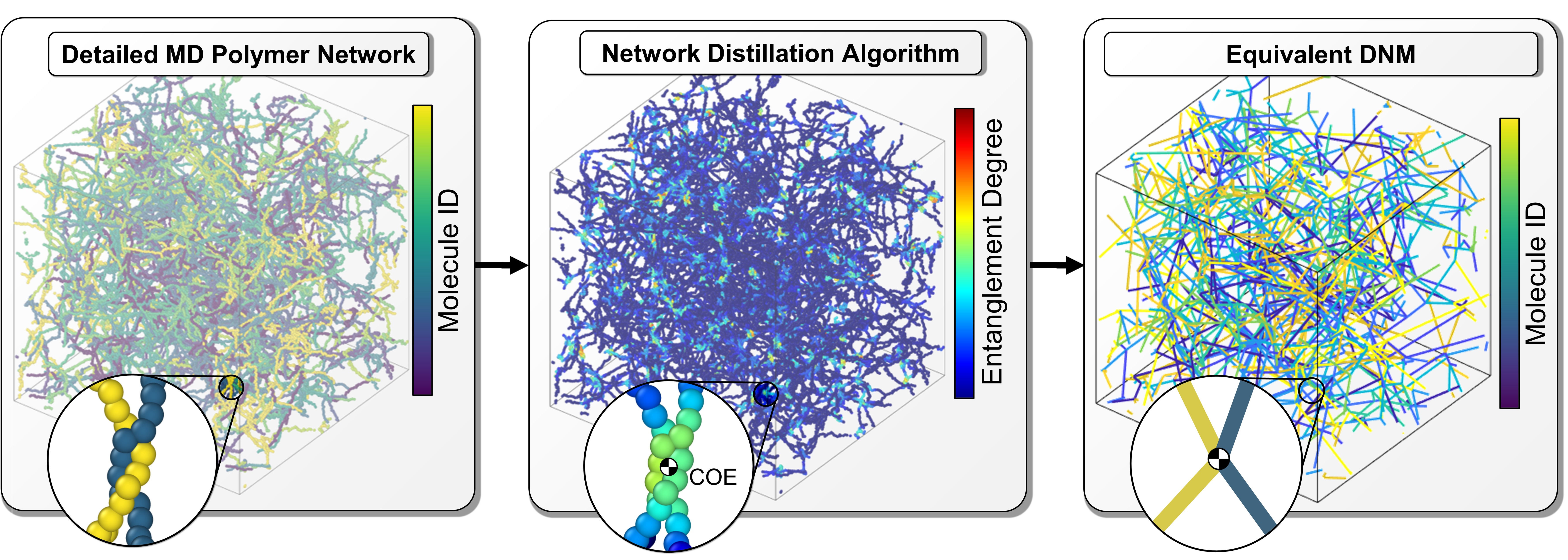}
        \caption{\label{fig:CGMD_to_DNM}
        \textbf{Primitive Path Distillation of Entangled Polymer Networks.} For any highly entangled polymer network modeled using CGMD (left), local entanglements may be algorithmically identified (center), so that the network may be distilled into a representative, reduced-order DNM (right).}
    \end{figure*}

Experimental studies have shown that entanglements can substantially enhance polymer mechanics \cite{edwards_effect_1986, tsukeshiba_effect_2005, klein_evidence_1978, de_rosa_effect_1994, ge_molecular_2013}. In melts, entanglements increase viscosity and govern reptation-controlled relaxation \cite{rubinstein_polymer_2003}. In crosslinked polymers, they can strongly enhance stiffness, strength, extensibility, and toughness by creating additional intermolecular load paths within the network \cite{mcleish_tube_2002,likhtman_quantitative_2002,gong_why_2010,sun_highly_2012,kamiyama_highly_2022,zheng_fracture_2022}. Experiments on double-network, tetra-gel, and topologically reinforced hydrogel architectures have further shown that load transfer through entanglements redistributes stress during deformation, thereby enhancing fatigue resistance and fracture resistance \cite{sakai_entanglement_2008, kamiyama_highly_2022, gong_why_2010, sun_highly_2012, creton_fracture_2016, zheng_fracture_2022}. For example, Kim et al. (2021) reduced the weight fraction of crosslinker in polyacrylamide hydrogels synthesized via free radical polymerization, producing networks in which entanglements greatly outnumbered crosslinks. They demonstrated that these gels exhibited up to three orders of magnitude higher toughness than those with predominantly crosslinks without sacrificing modulus or introducing hysteresis \cite{kim_fracture_2021}. Despite abundant experimental evidence supporting the benefits of entanglements, causally relating these micromechanical features to macroscopically observed properties remains extremely difficult.

Many experimental studies have explored synthesis-structure-property relationships in highly entangled polymers \cite{kim_fracture_2021,yuan_tough_2026,yang_polyacrylamide_2019,liu_polyacrylamide_2019,wang_polyacrylamide_2021,hassan_polyacrylamide_2022,kim_polyacrylamide_2022,wang_polyacrylamide_2023}, but experimental interrogation of entanglements remains challenging since they are nanoscale features that reside within the bulk of networks and are chemically identical to their surroundings. High-resolution imaging methods such as electron and atomic force microscopy require extensive sample preprocessing, do not image inside of networks, and cannot easily resolve 3D entanglement structures \cite{xu_thermosensitive_2022,martinez-garcia_beginners_2022}. Scattering methods such as small and wide-angle X-ray scattering can characterize mesh sizes and chain alignment but cannot easily distinguish chemical crosslinks from physical entanglements \cite{wang_trapped_2024,nishi_probe-saxs_2016}. As a result, relating micromechanics of entanglements to macroscale mechanical response relies heavily on physics-based models.

Theoretically, the mechanics of entangled polymers are studied using molecular theory \cite{de_gennes_reptation_1971}, physics-based constitutive models \cite{yang_hyperelastic_2025, vernerey_transient_2021, porter_entanglement_1966,edwards_statistical_1967,leygue_tube-based_2006}, and coarse-grained molecular dynamics (CGMD) \cite{zhang_polymer_2024, tejedor_molecular_2023, zou_molecular_2022}.  At macroscales, constitutive models based on molecular theory can predict the viscoelastic response of polymer melts and the hyperelastic stress--strain behavior of entangled networks \cite{edwards_effect_1986,wang_constitutive_2023,yang_hyperelastic_2025}. These mechanistic continuum models typically employ the primitive path assumption \cite{doi_theory_1988, doi_dynamics_1978} that the time-averaged entropic force acting through a polymer chain segment passes through the shortest end-to-end vector between the two topological constraints  -- crosslink or entanglement -- defining its end points. However, these continuum approaches cannot easily address discrete effects such as chain rupture \cite{shen_rate-dependent_2020,lamont_rate-dependent_2021,mulderrig_statistical_2023}, entanglement sliding \cite{hopmann_limitations_2020,liu_flow-driven_2025}, or damage delocalization \cite{deng_nonlocal_2023,masubuchi_review_2025}. Therefore, modeling approaches that explicitly track chain conformations and network configurations are needed.

One such approach is Kremer--Grest CGMD in which polymer chains are represented as beads connected by finitely extensible, nonlinear springs  (Fig.~\ref{fig:CGMD_to_DNM}, left and center) \cite{kremer_dynamics_1990}. This approach explicitly models excluded volume interactions between polymer chains so that entanglements emerge naturally \cite{hsu_primitive_2018,tejedor_molecular_2023} rather than being prescribed \textit{a priori} as in continuum approaches. However, Kremer-Grest and other CGMD models often require millions of beads with integration times on the order of picoseconds just to model representative volume elements (RVEs) of polymer networks that are $\sim$100~nm wide \cite{wagner_foundational_2025}. As a result, the spatiotemporal scales accessible to CGMD are limited to micrometers and microseconds so that quantitative structure-property mapping using CGMD remains intractable.

Towards overcoming this limitation, researchers have increasingly turned to discrete network models (DNMs) (Fig.~\ref{fig:CGMD_to_DNM}, right). In DNMs, only topological constraints (i.e., crosslinks \cite{sugimura_mechanical_2013,elbanna_dynamics_2013,alame_relative_2019,  wagner_network_2021,wagner_mesoscale_2022, wagner_mesoscale_2022,lei_network_2022,hartquist_fracture_2025} and -- if applicable -- entanglements \cite{gusev_molecular_2024,bernhard_phantom_2025,assadi_nonaffine_2025,huang_topological_2025}) are modeled explicitly as vertices. Meanwhile, the primitive paths between them are modeled as edges with implicit pairwise potentials based on the stretch-dependent Helmholtz free energy of entropic polymer chains. DNMs reduce computational cost proportionately to the molecular weight of the polymer mesh size, thereby granting access to spatiotemporal scales exceeding micrometers and microseconds \cite{wagner_foundational_2025} while preserving microstructural fidelity \cite{sugimura_mechanical_2013,wagner_network_2021,lei_network_2022,wagner_mesoscale_2022,wagner_coupled_2023,wagner_foundational_2025,hartquist_fracture_2025}. In the past few years, researchers have begun employing DNMs to model entangled polymers \cite{gusev_molecular_2024,bernhard_phantom_2025,assadi_nonaffine_2025} whereby entanglements are treated as vertices across which neighboring chain segments can redistribute their effective length. 

Recently, Assadi et al. (2025) showcased the utility of entangled DNMs for predicting nonlinear rheological behavior of melts and gelation fractions of crosslinkers in polymer networks \cite{assadi_nonaffine_2025}. Huang et al. (2025) used a similar approach to elucidate how entanglements delocalize crack-tip stresses via sliding in elastomers \cite{huang_topological_2025}. Gusev and Bernhard (2024, 2025) have demonstrated the significant computational cost savings afforded by using DNMs (in lieu of CGMD) to accurately predict storage and loss moduli of PDMS \cite{gusev_molecular_2024,bernhard_phantom_2025}. Given these benefits, bottom-up, physics-based approaches that accurately distill entangled CGMD networks into their equivalent DNM representations would be of great value to predictive, multiscale structure-property mapping efforts. 

To address this need, we introduce a quantitative method of identifying inter- and intra-chain entanglements along the backbone of discretely modeled polymer chains based on the Gaussian Linking Number (GLN). Whereas traditional GLN-based approaches output a single value of linking number for any pair of chains, our approach defines the entire set of entanglements occurring along the backbone of a polymer. This is critical for polymer networks, which often include multiple independent entanglements between the same two chains, as well as intrachain entanglements. Our approach also introduces a geometric center of entanglement (COE) that defines the load-transmission position between chain segments. We validate this approach using Kremer-Grest simulations before leveraging it to introduce a physically-motivated distillation procedure that converts CGMD networks into mechanically predictive DNMs (Fig.~\ref{fig:CGMD_to_DNM}). 

In the remainder of this work, we detail our procedure for conducting entangled Kremer-Grest simulations (\textbf{Section~\ref{sec:ModelingEntangledChains}}), define our methods of characterizing entanglement (\textbf{Section~\ref{sec:CharacterizingPolymerEntanglement}}), and then introduce our CGMD-to-DNM distillation procedure (\textbf{Section~\ref{sec:DistillingPolymerNetworks}}). We then physically justify the premise of our distillation procedure by exploring the mechanical relevance (\textbf{Section~\ref{sec:ReliabilityOfPrimitivePaths}}) and mobility (\textbf{Section~\ref{sec:ReliabilityOfPrimitivePaths}}) of the newly introduced COEs. Finally, we verify our distillation procedure's ability to generate DNMs that accurately reproduce the mechanical stress predictions and network structures of CGMD models \textbf{Section~\ref{sec:DNM_Mechanical_Response}}.

%%%%%%%%%%%%%%%%%%%%%%%%%%%%%%%%%%%%%%%%%%%%%%%%%%%%%%%%%%%%%%%%%%%%%%%%%%%%%%%%%%%%%%%%%%%%%%%%%%%%
\section{\label{sec:Methods}METHODS}%:\protect\\

In this section, we detail the initiation and simulation procedures for our Kremer-Grest models; introduce the GLN and COE used to characterize entanglement degrees and positions, respectively; and outline the CGMD-to-DNM distillation procedure.

%%%%%%%%%%%%%%%%%%%%%%%%%%%%%%%%%%%%%%%%%%%%%%%%%%%%%%%%%%%%%%%%%%%%%%%%%%%%%%%%%%%%%%%%%%%%%%%%%%%%
\subsection{\label{sec:ModelingEntangledChains}MODELING ENTANGLED CHAINS}%:\protect\\ 

To run Kremer-Grest simulations, we first initiated the positions of $\mathcal{N}$ polymer chains using an on-lattice, self–avoiding walk (SAW) algorithm \cite{kremer_dynamics_1990,de_gennes_scaling_1979}. The chains were grown inside of a 3D periodic domain on a cubic lattice with a separation scale of one Kuhn length, $b$. The length of each chain, $Nb$, was prescribed by assigning its number of growth steps, $N$ (i.e., Kuhn segments). The number of lattice sites per unit box edge was set according to $n=\lceil (N_{tot}/\phi)^{1/3} \rceil$, where $N_{tot}$ is the total number of beads for all chains, $\phi$ is the desired fraction of sites occupied by polymer beads, and $\lceil \square \rceil$ denotes rounding up to the nearest integer to ensure exact lattice spacing of $b$ across the periodic boundaries. To improve the statistical likelihood of entanglement between chains, we generally set $\phi\geq 0.1$. However, setting $\phi>0.5$ significantly slowed the SAW algorithm due to increased scarcity of vacant growth sites. As such, we set $\phi=0.1$ and $\phi=0.5$ for the two-chain and networked systems of \textbf{Sections~\ref{sec:ReliabilityOfPrimitivePaths}-\ref{sec:COEDiffusion}} and \textbf{\ref{sec:DNM_Mechanical_Response}}, respectively.

Initiation sites for each chain were randomly selected from the available lattice sites. Growth steps were carried out by alternating between all chains so that polymerization of each chain was effectively concurrent. Occupied lattice sites were removed from the pool of available growth sites following standard SAW procedure. Once all chains were initiated, we adjusted the chain stretch by applying an affine, isotropic expansion to every bead following the deformation tensor, $\bm F = \lambda_I \bm I$. Here, $\bm I$ is the identity tensor and $\lambda_I$ is an isotropic stretch set according to \textbf{Sections~\ref{sec:ReliabilityOfPrimitivePaths}-\ref{sec:COEDiffusion}} and \textbf{\ref{sec:DNM_Mechanical_Response}} for two-chain systems and networks, respectively.  

To relax the chains and sample their conformational states, we ran NVT Brownian Dynamics simulations in LAMMPS \cite{thompson_lammps_2022}. The position, $\bm x_i$, of the $i^{th}$ bead was updated in time, $t$, according to:

\begin{equation}
    \gamma  \frac{d \bm x_i}{dt} =  \sum_j \bm f_{ij} + \sqrt{2 \gamma k_B T} \bm \eta_i.
    \label{eq:position}
\end{equation}

\noindent Here $\gamma$ is a damping coefficient capturing viscous drag, $\sum_j \bm f_{ij}$ is the unbalanced force acting on each bead due to all pairwise interactions, $k_B T$ is the thermal energy, $k_B$ is the Boltzmann constant, $T$ is the ambient temperature, and $\bm \eta_i$ is a stochastic vector introducing thermal forces \cite{doyle_brownian_2005}. Values of $\bm \eta_i$ are sampled from a Gaussian distribution with mean $\bm 0$ and variance $dt^{-1}$ using a delta-correlated stationary process so that there is no correlation between $\bm \eta_i(t)$ and $\bm \eta_i(t+dt)$ \cite{doi_theory_1988, kremer_dynamics_1990}. 

Unbalanced forces consist of bonded interactions that represent chemical bonds along the polymer backbone, and non-bonded excluded volume interactions between all beads within cutoff distance $b$ of each other. Bonded forces are captured as $-\partial U_b/\partial \bm r_{ij}$ where the interaction potential is defined by LAMMPS's built-in nonlinear, finitely extensible relation \cite{rector_simulation_1994}:

\begin{equation}
    U_b = \frac{E (r_{ij} - b)^2}{\ell^2 - (r_{ij} - b)^2}.
    \label{eq:FENE}
\end{equation}

\noindent Here $E$ is an energy scale set sufficiently large to approximate rigid bonds \cite{wagner_foundational_2025}, $r_{ij}$ is the instantaneous bond length between attached beads $i$ and $j$, and $\ell$ specifies the finite length by which the bond may deviate from equilibrium length $b$. Similarly, pairwise excluded volume interactions are captured by $-\partial U_{WCA}/\partial \bm r_{ij}$ where $U_{WCA}$ is the purely repulsive Weeks--Chandler--Andersen (WCA) potential \cite{weeks_role_1971} given by:

\begin{equation}
    U_{\mathrm{WCA}} =
    \begin{cases}
        4\epsilon \left[ \left(\dfrac{\sigma}{d_{ij}}\right)^{12} - \left(\dfrac{\sigma}{d_{ij}}\right)^{6} \right], & d_{ij} \leq b, \\
        0, & d_{ij} > b
    \end{cases}.
    \label{eq: WCA}
\end{equation}

\noindent Here $\epsilon$ is an energy scale that modulates the interaction stiffness, and $\sigma$ is a characteristic interaction length set to $b/2^{1/6}$ so that force goes to zero when the pairwise distance $d_{ij}$ goes to $b$. The value of $\sigma$ and the cutoff criterion for this potential enforce that there are only short-range repulsive interactions and no long-range attraction between non-bonded pairs consistent with good solvent conditions \cite{merlitz_theoretical_2016,wagner_mesoscale_2022}. For a list of fundamental model constants and parameters, see \textbf{Tables~\ref{tab:fundamental parameters}} and \textbf{\ref{tab:secondary parameters}}, respectively.

All simulations were performed in the reduced Lennard--Jones units of LAMMPS. The fundamental energy, length, and time scales are set by the thermal energy $k_B T$, the Kuhn length $b$, and the characteristic bead diffusion time $\tau_0$, respectively. For generality, we adopted $b = 1$~nm, which reasonably mirrors flexible polymers such as polyacrylamide ($b = 1.6$~nm) and polyethylene glycol ($b = 0.7$~nm) \cite{rubinstein_polymer_2003,lee_molecular_2008}. The corresponding diffusion timescale was set to $\tau_0 = 10^{-8}$ s, consistent with the timescales required for nanometer-scale beads to diffuse their own characteristic size in aqueous solutions at ambient temperatures \cite{wagner_foundational_2025}. Derived parameters are defined accordingly: the diffusion coefficient was computed as $D = b^2/\tau_0$, yielding a damping coefficient $\gamma = k_B T / D$ \cite{einstein_uber_1905}. The timestep was set to $dt = 10^{-5}\tau_0$, which was the largest value found to maintain numerical stability.

\begin{table}[!htbp]
\caption{\label{tab:fundamental parameters}%
Fundamental model parameters and constants.
}
\begin{ruledtabular}
\begin{tabular}{lcccl}
\textrm{Description} &
\textrm{Symbol} &
\makecell{Model \\ Value} &
\makecell{SI \\ Value} &
\textrm{Units} \\
\colrule
Boltzmann Constant   & $k_B$ & 1 & $1.38\times10^{-23}$ & J/K \\
Temperature        & $T$   & 1 & 293 & K \\
Kuhn Length        & $b$   & 1 & $1.0\times10^{-9}$ & m \\
Diffusion Timescale& $\tau_0$ & 1 & $1.0\times10^{-8}$ & s \\
\end{tabular}
\end{ruledtabular}
\end{table}

\begin{table}[!htbp]
\caption{\label{tab:secondary parameters}Secondary model parameters.}
\begin{ruledtabular}
\begin{tabular}{lccl}
\textrm{Description} &
\textrm{Symbol} &
\textrm{Value} &
\begin{tabular}{c}Reduced\\Units\end{tabular} \\
\colrule
Diffusion Coefficient & $D$ & $1$ & $b^2/\tau_0$ \\
Damping Coefficient & $\gamma$ & $1$ & $k_B T/D$ \\
Time Step & $dt$ & $10^{-5}$ & $\tau_0$ \\
WCA Energy Scale & $\epsilon$ & $1$ & $k_B T$ \\
WCA Length Scale & $\sigma$ & $2^{-1/6}$ & $b$ \\
Bond Energy Scale & $E$ & $800$ & $k_B T$ \\
Bond Extensibility Length & $\ell$ & $1$ & $b$ \\
Lattice Fill Ratio & $\phi$ & 0.1--0.5 & -\\
\end{tabular}
\end{ruledtabular}
\end{table}

%%%%%%%%%%%%%%%%%%%%%%%%%%%%%%%%%%%%%%%%%%%%%%%%%%%%%%%%%%%%%%%%%%%%%%%%%%%%%%%%%%%%%%%%%%%%%%%%%%%%
\subsection{\label{sec:CharacterizingPolymerEntanglement}CHARACTERIZING PAIRWISE ENTANGLEMENTS}

\begin{figure}[!htbp]
    \includegraphics[width=\columnwidth]{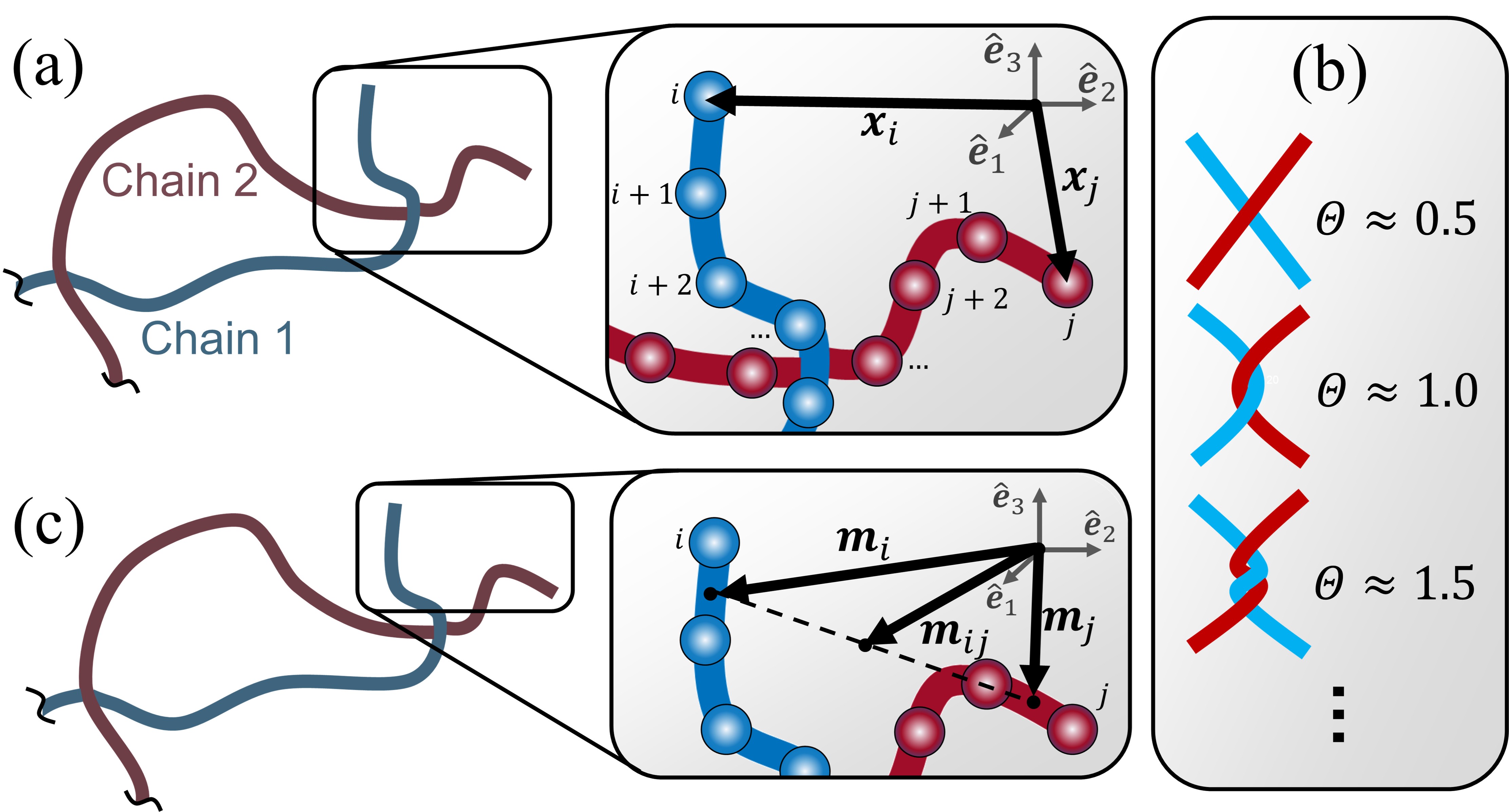}
        \caption{\label{fig:Linking_Number_Figure} \textbf{Characterizing Entanglements. (a)} Vector convention for computing $\varTheta$. \textbf{(b)} Interpretation of the Gaussian linking number, $\varTheta$. \textbf{(c)} Vector convention for computing $\bm x_c$.}
    \end{figure}

To characterize the pairwise entanglement state between any two polymer chains, we employ the GLN, $\varTheta\in (-\infty,\infty)$. For any set of two open curves (e.g., chains 1 and 2 in Fig.~\ref{fig:Linking_Number_Figure}(a)), $\varTheta$ indicates how many times one curve wraps around the other. For example, $\varTheta=0$ corresponds to parallel chains (i.e., no wrapping), $\varTheta=0.5$ indicates a half-wrap (i.e., crossing without wrapping), $\varTheta=1$ indicates a single wrap, and so on (Fig.~\ref{fig:Linking_Number_Figure}(b)). The sign of $\varTheta$ simply encodes the handedness of the winding (i.e., clockwise or counterclockwise). 

For the CGMD polymers initiated in \textbf{Section~\ref{sec:ModelingEntangledChains}}, whose bead positions are at $\bm x_i$ and $\bm x_j$ for chains 1 and 2, respectively (Fig.~\ref{fig:Linking_Number_Figure}(a)), the GLN may be computed as:

\begin{equation}
    \varTheta=\frac{1}{4\pi}\sum_{i=1}^{N_1}\sum_{j=1}^{N_2}\vartheta_{ij}.
    \label{eq:LN}
\end{equation}

\noindent Here $i \in [1,N_1+1]$ and $j \in [1,N_2+1]$ denote the indices of beads along chains 1 and 2 while $N_1$ and $N_2$ indicate the number of Kuhn segments in each chain, respectively. The scalar value $\vartheta_{ij}\in(-\infty,\infty)$ represents how much each pair of segments from chains 1 and 2 contribute to the overall GLN. Recently, Ahmed et al. (2020) introduced a method of computing $\vartheta_{ij}$ for entangled chains in glassy polymers that yields GLN values invariant to both chain length and discretized step size, $b$ \cite{ahmad_characterization_2020}. This method provides the pairwise segmental contributions as:

    \begin{equation}
        \vartheta_{ij} = \Omega(\bm s,\bm t,\bm u) + \Omega(\bm u,\bm v,\bm s),
        \label{eq:vartheta_klmn}
    \end{equation}

\noindent where $\bm s = \bm x_i - \bm x_j$, $\bm t = \bm x_{i+1} - \bm x_j$, $\bm u = \bm x_{i+1} - \bm x_{j+1}$, $\bm v = \bm x_i - \bm x_{j+1}$, and $\Omega (\square)$ denotes the discrete solid-angle operation given by \cite{ahmad_characterization_2020}:

    \begin{equation}
        \begin{aligned}
            \Omega&(\bm s,\bm t,\bm u) = ... {} \\
            &2 \arctan \left(\frac{\bm s \cdot (\bm t \times \bm u)}{stu + (\bm s \cdot \bm t)u + (\bm t \cdot \bm u)s + (\bm u \cdot \bm s)t} \right).
        \end{aligned}
        \label{eq:solid_angle}
    \end{equation}

    % fig 3
      
While the GLN of Eqs.~(\ref{eq:LN})--(\ref{eq:solid_angle}) characterizes the degree of entanglement between two polymer chains, we must also understand the effective entanglement position between them to assess the primitive path assumption and achieve DNM distillation. To do so, we define the geometric COE as the weighted centroid position of all segment–pair midpoints according to:

\begin{equation}
    \bm x_c = \frac{\sum_{i,j} \vartheta_{ij}^p \,\bm{m}_{ij}}{\sum_{i,j} \vartheta_{ij}^p},
    \label{eq:COE}
\end{equation}

\noindent where $\bm m_{ij}=(\bm m_i+\bm m_j)/2$ is the midpoint between beads $i$ and $j$ of chains 1 and 2, respectively (Fig.~\ref{fig:Linking_Number_Figure}(c)) \cite{karayiannis_structure_2009}. The power $p$ is used to modulate sensitivity of $\bm x_c$ to $\vartheta_{ij}$, such that higher values of $p$ increase the influence of $(i,j)$ pairs with smaller pairwise separation distances. We find that setting $p=2$ produces COE predictions that closely align with primitive path force directions as explored in \textbf{Section~\ref{sec:ReliabilityOfPrimitivePaths}}. 

\begin{figure}[!htbp]
    \includegraphics[width=\columnwidth]{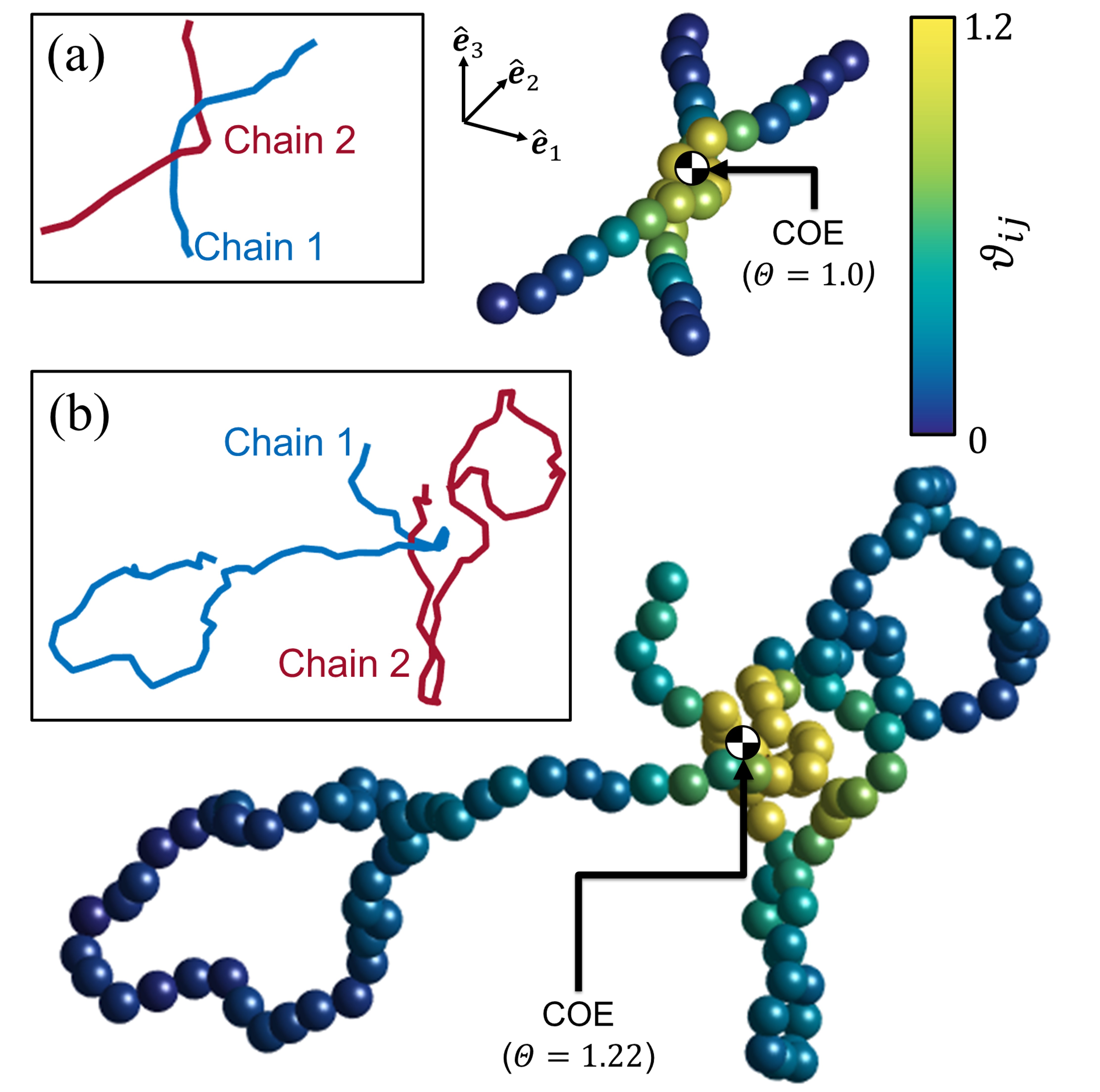}
        \caption{\label{fig:COEs} \textbf{Visualizing the COE. (a)} Benchmark case in which $N_1=N_2=20$, all chain ends are in $xy$-plane, and $\varTheta\approx 1$. \textbf{(b)} Stochastically generated system with $N_1=62$ and $N_2=60$ for which the COE is less intuitive. The color map in both panels denotes the local weighting, $\vartheta_{ij}$ of each segment to the overall linking number. For visual clarity, the red and blue curves of the insets illustrate the distinct paths of chains 1 and 2, respectively.}
    \end{figure}

Fig.~\ref{fig:COEs}(a) illustrates the COE for a benchmark problem in which all four chains' ends are co-planar, $\varTheta\approx 1$, and chain stretch is high, so that the COE is simply the approximate position where the chains intersect. However, Fig.~\ref{fig:COEs}(b) illustrates a stochastically generated pair of chains for which the COE is not obvious but becomes quantitatively predictable via Eq.~\eqref{eq:COE}. We evaluate the efficacy of Eq.~\eqref{eq:COE} at predicting the effective locations of inter-chain load transmission in \textbf{Section~\ref{sec:ReliabilityOfPrimitivePaths}}.

    % fig 4

%%%%%%%%%%%%%%%%%%%%%%%%%%%%%%%%%%%%%%%%%%%%%%%%%%%%%%%%%%%%%%%%%%%%%%%%%%%%%%%%%%%%%%%%%%%%%%%%%%%%
\subsection{\label{sec:DistillingPolymerNetworks}DISTILLING ENTANGLED NETWORKS}

\begin{figure}[!htbp]
    \includegraphics[width=\columnwidth]{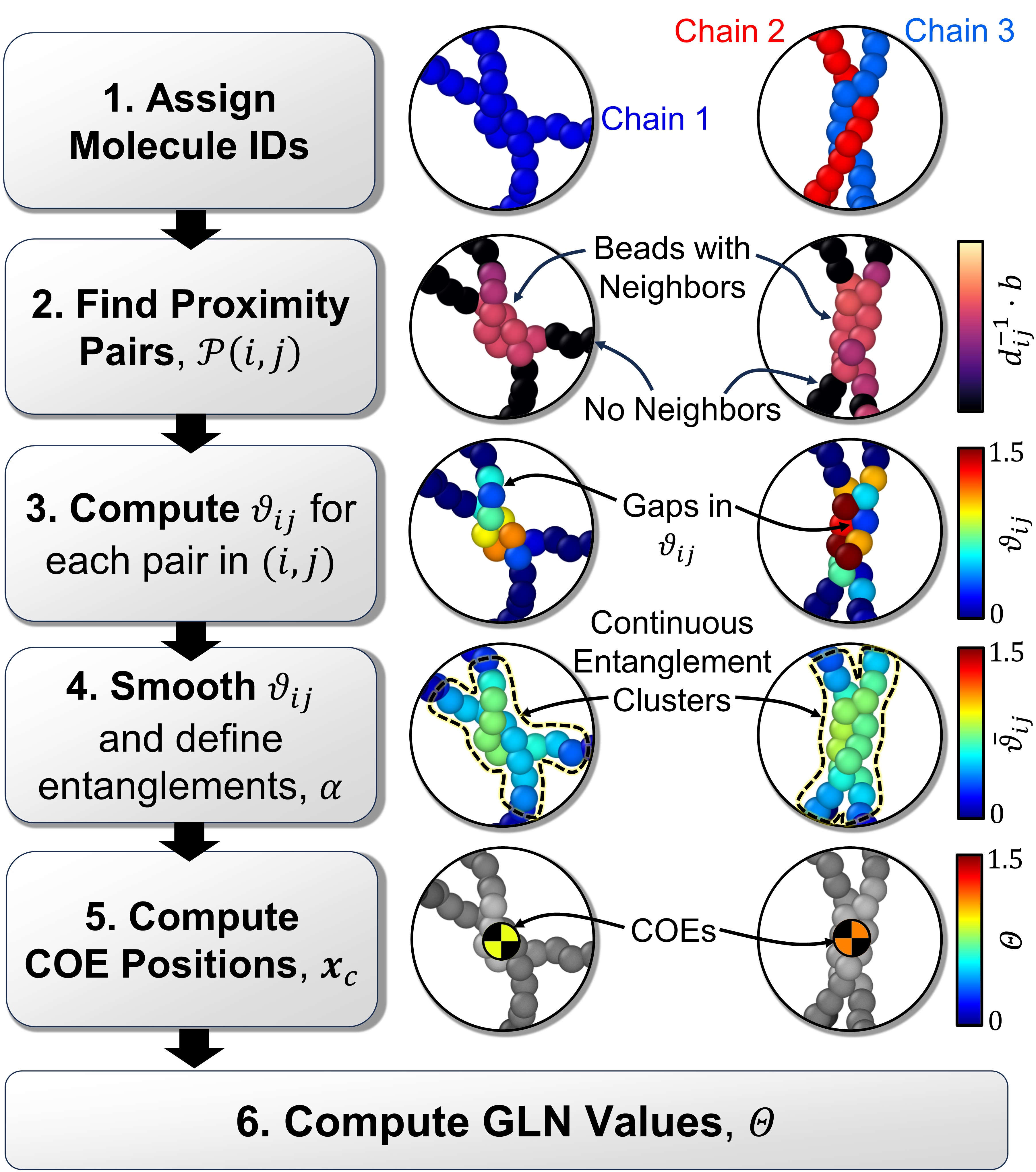}
        \caption{\label{fig:Local_Entanglement} \textbf{} 
        \textbf{Defining local entanglements.} The procedural algorithm for identifying local entanglements along the backbone of chains is outlined, along with visualizations for each step from actual CGMD networks. Instances of both intramolecular (left) and intermolecular (right) entanglements are depicted. Step 1: Define CGMD network and assign molecule numbers to each chain. Step 2: Identify set of proximity pairs, $\mathcal{P}(i,j)$, within distance $d_c$ of one another. Color bar represents normalized inverse distance to closest nearest neighbor, $b/d_{ij}$. Step 3: Compute $\vartheta_{ij}$ for all segment pairs $i \rightarrow i+1$ and $j \rightarrow j+1$. Step 4: Smooth $\vartheta_{ij}$ over a window of length $\ell_w$.  Steps 5-6: Compute $\bm x_c$ and $\varTheta$ for each entanglement using Eqs.~\eqref{eq:LN}-\eqref{eq:COE}.
        }
    \end{figure}

\begin{figure}[!htbp]
    \includegraphics[width=\columnwidth]{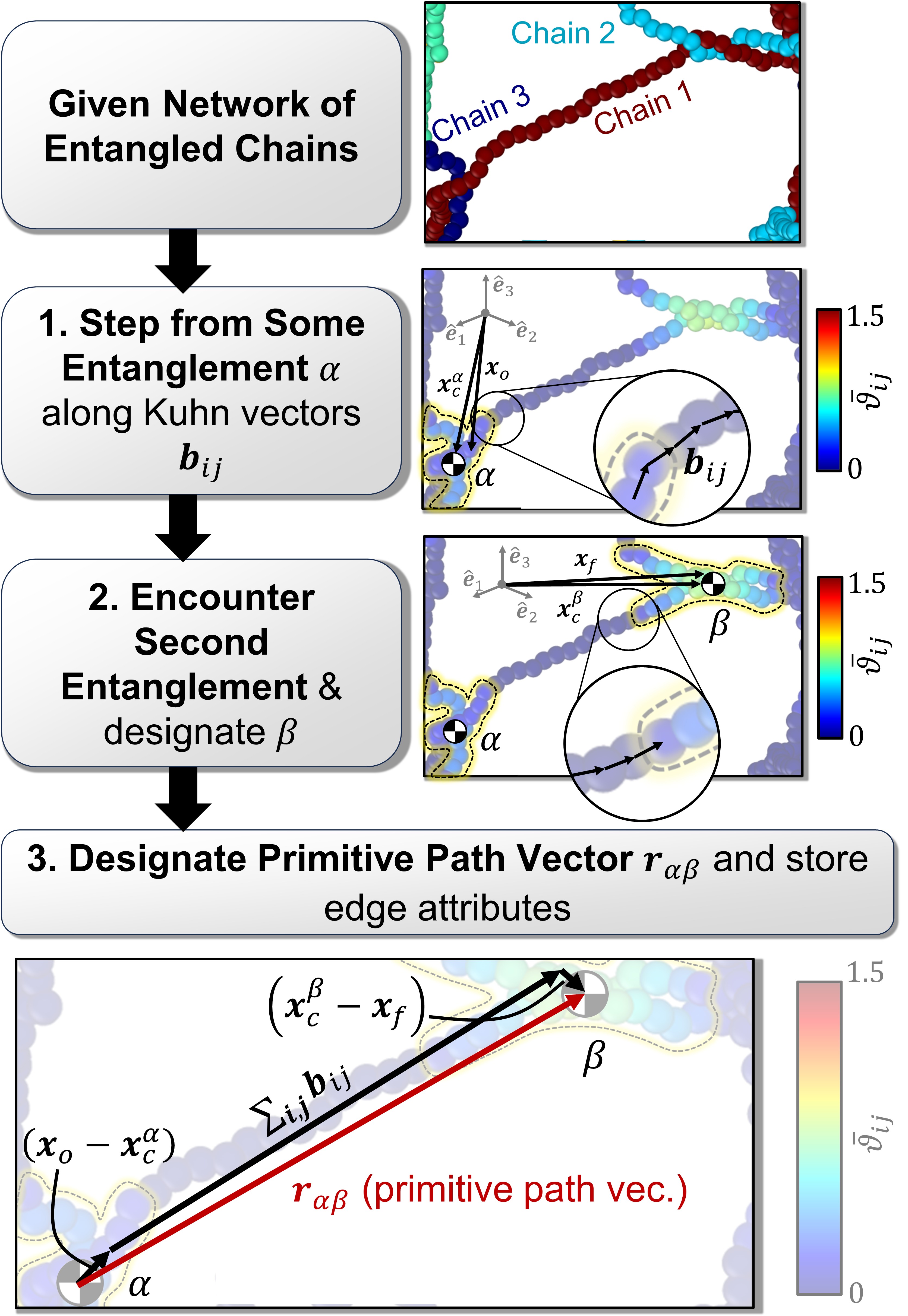}
        \caption{\label{fig:Primitive_Paths} 
        \textbf{Defining primitive paths.} The procedural algorithm for identifying primitive paths between entanglements is shown. Given a network of indexed chains and entanglement clusters (top), primitive paths may be identified using the following steps. Step 1: Beginning from some randomly selected entanglement, $\alpha$, step its bead closes to $\bm x_c^\alpha$ (at position $\bm x_o$) along the Kuhn vectors, $\bm b_{ij}$. Step 2: Step along the polymer until a bead belonging to another entanglement, $\beta$ is encountered and continue until reaching the bead closest to $\bm x_c^\beta$ (at positions $\bm x_f$). Step 3: Define the primitive path vector between entanglements $\alpha$ and $\beta$ as $\bm r_{\alpha \beta} = (\bm x_o - \bm x^\alpha_c) + \sum_{i,j} \bm b_{ij} + (\bm x^\beta_c - \bm x_f)$.}  
    \end{figure}

While the entanglements defined in \textbf{Section~\ref{sec:CharacterizingPolymerEntanglement}} characterize the overall wrapping state between any two open chains, in a real polymer network the same two polymer chains can entangle in multiple places along their backbone. This gives rise to multiple distinct clusters of entangled beads that transmit interchain loads and which must be accounted for in mechanically predictive DNMs. Therefore, to distill CGMD networks into representative DNMs, we here introduce an algorithm that first identifies and characterizes these ``local'' entanglement clusters following the procedure of Fig.~\ref{fig:Local_Entanglement}. It then defines the primitive paths between these entanglement clusters following the procedure of Fig.~\ref{fig:Primitive_Paths}. These steps are detailed as follows.
 
\textbf{Defining local entanglements:} Physical contact is required for load transmission between entangled chains. Therefore, to identify load transmitting entanglement clusters in CGMD networks, we first perform a pairwise distance check between all beads to identify potential entanglement contributors. For each bead $i$, all neighbors $j$ within cutoff distance $d_c$ are added to a pairs list, $\mathcal{P}(i,j)$, provided they are not on the same chain within a contour distance, $\ell_s$, of each other along the backbone (Fig.~\ref{fig:Local_Entanglement}, steps 1-2). Given our bead diameter of $b$, we found that the algorithm nicely captures physically relevant contacts if $d_c/b=1.5$, even in the presence of thermal oscillations. Since chains' lengths must significantly exceed their persistence lengths, $\ell_p$, to form intramolecular entanglements or knots, we set the intramolecular exclusion distance to $\ell_s\sim 2\ell_p \sim 4b$ (see \textbf{Supplementary Information (SI) Section S1} for details) \cite{flory_statistical_1969, wagner_treadmilling_2021}. Single-chain segments less than this length cannot form intramolecular entanglements and thus pairwise distance checks are not needed for beads comprising the same chain separated by less than this contour length.

    % fig 5
    
Once we have identified the set of proximity pairs, $\mathcal{P}(i,j)$, we must determine which beads in this set contribute to entanglements, and which contributing beads work together to form distinct entanglement clusters. Note that in the remainder of this work, we denote CGMD beads by ``$i$'' and their pairwise relations by ``$ij$'', while we denote entanglements by ``$\alpha$'' and their pairwise relations by ``$\alpha \beta$''. To identify entanglement-contributing beads, we compute $\vartheta_{ij}$ for every $(i,j)$ pair using Eq.~\eqref{eq:vartheta_klmn}. If $|\vartheta_{ij}|>10^{-12}$, then the contribution of beads $i$ and $j$ to an entanglement is treated as non-negligible. Next, to identify which beads with finite $\vartheta_{ij}$ contribute to a distinct entanglement, $\alpha$, we average $|\vartheta_{ij}|$ over a moving window (Fig.~\ref{fig:Local_Entanglement}, step 3) centered at bead $i$ and traversing a distance of $\ell_w$ along the backbone of the chain to which $i$ belongs. For discrete chains, this amounts to averaging over the window $i\pm \lfloor N_w/2 \rfloor$ where $N_w = \lceil \ell_w/b \rceil+1$ is the number of beads comprising length $\ell_w$. We set $\ell_w=2\ell_p$, which is large enough to close gaps of non-finite $|\vartheta_{ij}|$ residing within entanglement clusters, but small enough to prevent merging of distinct neighboring entanglements.  This process produces smoothed values of $\vartheta_{ij}$ along the backbones of chains (Fig.~\ref{fig:Local_Entanglement}, steps 3-4).  

With $\vartheta_{ij}$ smoothly defined, distinct sections of chains belonging to entanglements, $\alpha$, are defined as the set, $\mathcal{S}_\alpha$, of contiguous beads with finite values of $\vartheta_{ij}$ along each chain. Two contiguous sets of beads with finite $\vartheta_{ij}$ (denoted $\mathcal{S}_\alpha$ and $\mathcal{S}_\beta$) are considered to belong to a single entanglement cluster (denoted $\alpha$) if any beads $i\in \mathcal{S}_\alpha$ and $j\in \mathcal{S}_\beta$ belong to the set of proximity pairs $\mathcal{P}(i,j)$ from step 1. This merging is applied iteratively until all $\mathcal{P}(i,j)$ pairs belong to a single entanglement cluster (Fig.~\ref{fig:Local_Entanglement}, step 4). Finally, the COE position, $\bm x_c^\alpha$, and GLN, $\varTheta_\alpha$, of each entanglement are computed using Eq.~\eqref{eq:COE} and Eqs.~\eqref{eq:LN}, respectively. However, $\vartheta_{ij}$ and $\bm{m}_{ij}$ are now evaluated only for segment pairs belonging locally to entanglement $\alpha$ (Fig.~\ref{fig:Local_Entanglement}, steps 5-6). Both $\bm x_c^\alpha$ and $\varTheta_\alpha$ are stored as vertex attributes for each entanglement in distilled DNMs.

    % fig 6
    
\textbf{Defining primitive paths:} To identify the primitive paths between two attached entanglements, $\alpha$ and $\beta$ (Fig.~\ref{fig:Primitive_Paths}, top row), we select the bead belonging to $\mathcal{S}_\alpha$ closest to the COE position, $\bm{x}_c^\alpha$, and designate its position as $\bm x_o$ (Fig.~\ref{fig:Primitive_Paths}, step 1). We then step from $\bm x_o$ along the bonded neighbors of the CGMD polymer backbone until encountering a bead belonging to a different entanglement, $\beta$. We continue stepping until reaching the bead belonging to $\beta$ closest to $\bm x_c^\beta$ whose position we designate $\bm x_f$ (Fig.~\ref{fig:Primitive_Paths}, step 2). The primitive path vector between entanglements $\alpha$ and $\beta$ is then defined as
$\bm r_{\alpha \beta} = (\bm x_o - \bm x_c^\alpha) + \sum_{i,j} \bm b_{ij} + (\bm x_c^\beta - \bm x_f)$,
where $\bm x_c^\alpha$ and $\bm x_c^\beta$ are the COE positions of entanglements $\alpha$ and $\beta$, respectively, and $\sum_{i,j} \bm b_{ij}$ is the sum of Kuhn vectors along the traversed backbone segment (Fig.~\ref{fig:Primitive_Paths}, step 3). This stepwise construction is carried out until beads have been swept over and ensures that all values of $\bm r_{\alpha \beta}$ are consistent with periodic boundary conditions.

Once each primitive path, $\bm r_{\alpha \beta}$, is defined, $\alpha$ and $\beta$ are stored in a bonded pairs list, $\mathcal{E}(\alpha,\beta)$, for DNM representation. Additionally, the primitive path end-to-end vector $\bm r_{\alpha \beta}$, and primitive path length (i.e., number of primitive path segments, $N_p$), are stored as edge attributes in the DNM. While specific to linear chains, this step-growth method can be slightly adjusted to accommodate branched and crosslinked chain topologies \cite{masubuchi_primitive_2006, davidson_nonaffine_2016}.

%%%%%%%%%%%%%%%%%%%%%%%%%%%%%%%%%%%%%%%%%%%%%%%%%%%%%%%%%%%%%%%%%%%%%%%%%%%%%%%%%%%%%%%%%%%%%%%%%%%%
\section{\label{sec:ReliabilityOfPrimitivePaths}COES PREDICT RELIABLE LOAD TRANSMISSION AND PRIMITIVE PATHS}

\begin{figure}[!htbp]
        \includegraphics[width=\columnwidth]{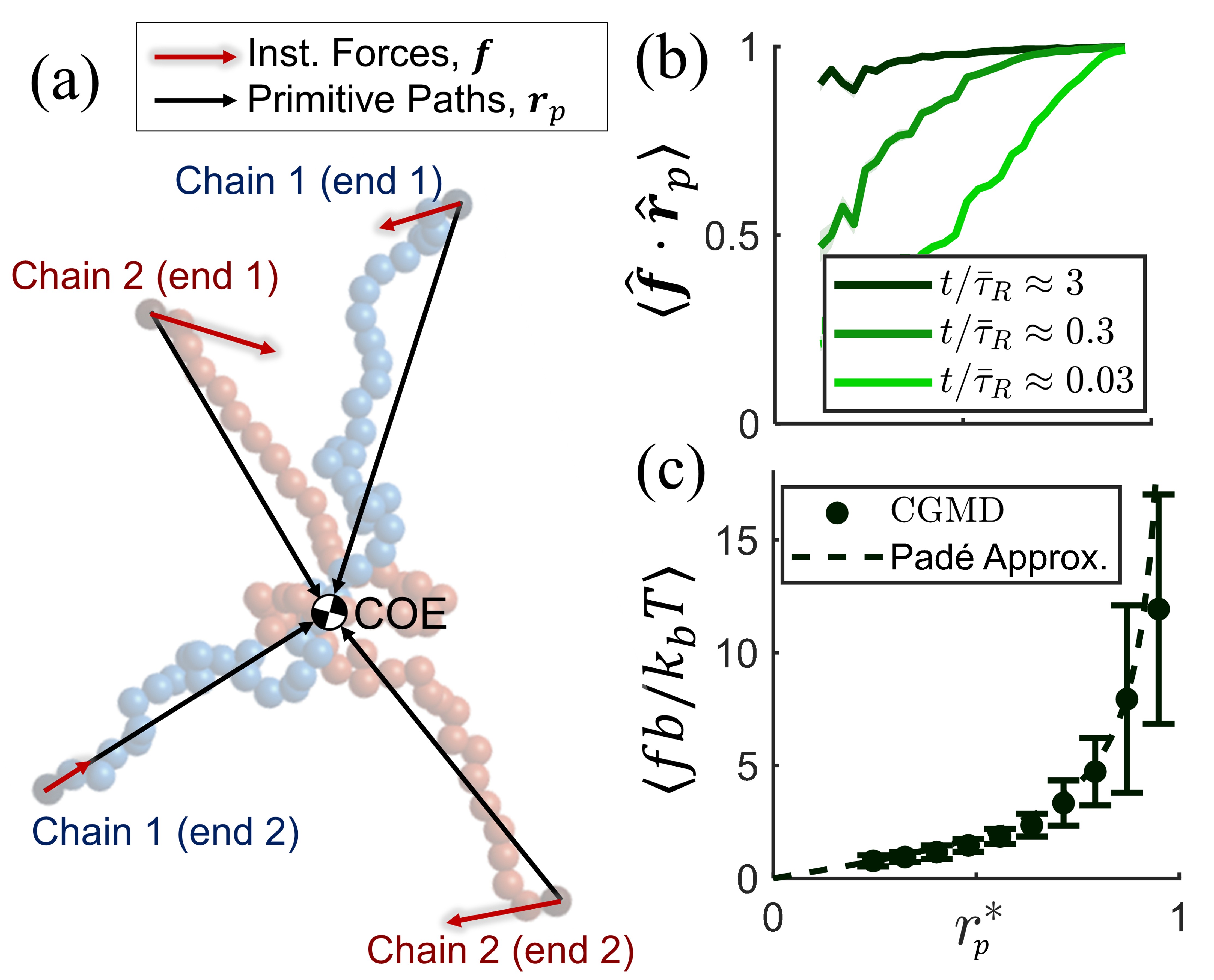}
        \caption{\label{fig:Primitive_Path_Alignment}
        \textbf{Primitive path alignment of entropic forces.} 
        \textbf{(a)} Instantaneous end forces $\bm{f}$ (red) and primitive path vectors $\bm{r}_{p}$ (black) referenced from the COE for an entangled pair of chains.
        \textbf{(b)} Ensemble-averaged alignment $\langle \hat{\bm f} \cdot \hat{\bm r}_p  \rangle$ versus normalized stretch $r^*$ for different holding times $t/\bar{\tau}_R$.
        \textbf{(c)} Force–stretch relation, $\langle fb/k_B T\rangle$ versus $r_p^*$, comparing ensemble averages (symbols) with the freely jointed chain prediction from Eq.~\eqref{eq:Pade} (dashed line).}
    \end{figure}

Primitive path analysis posits that the forces acting through the ends of a polymer chain segment (red arrows in Fig.~\ref{fig:Primitive_Path_Alignment}(a)) time-averaged over durations exceeding the longest relaxation time (i.e., the Rouse time \cite{rubinstein_polymer_2003}) will be directed along the end-to-end vector (black arrows in Fig.~\ref{fig:Primitive_Path_Alignment}(a)) between the two topological end-points that define it, which can be crosslinks or entanglements. Statistical mechanics dictates that these time-averaged forces are entropically governed tensile forces whose magnitudes are given by $\partial \psi /\partial r_p$, where $\psi$ is the Helmholtz free energy of the chain segment \cite{flory_molecular_1985}. For the freely jointed, finitely extensible chains of the Kremer-Grest model, the Helmholtz free energy of a primitive path segment should be well represented by the Pad\'e approximation \cite{cohen_pade_1991} of the finitely extensible Helmholtz free energy introduced by \cite{kuhn_statistical_1946}:

\begin{equation}
    \psi = k_B T \left[ \frac{\lambda_p^2}{2} - N_p \ln (N_p-\lambda_p^2) \right],
    \label{eq:Pade}
\end{equation}

\noindent where $\lambda_p = r_p/\sqrt {N_p} b$ is the stretch of a primitive path with end-to-end vector $\bm r_p$ composed of $N_p$ Kuhn segments. We hypothesize that for entangled chains, the time-averaged entropic forces will pass approximately through the COE introduced through Eq.~\eqref{eq:COE}, regardless of chain stretch and contour length. 

To explore these primitive path assumptions and test our hypothesis, we carried out $n=25{,}000$, two-chain Kremer-Grest simulations in which inter-chain entanglements emerged naturally. Chains were initiated at a packing fraction of $\phi=0.1$, with various lengths and end-to-end stretches following the procedure of \textbf{Section~\ref{sec:ModelingEntangledChains}}. 

Chain lengths were prescribed by randomly selecting the number of Kuhn segments in each chain from the uniform integer distribution in the range $N\in[20,100]$. We found that when chains had $N<20$ segments, they rarely formed entanglements with $|\varTheta_\alpha|>1$ (i.e., at least one full wrap), in which case they often disentangled while sampling entropic force-fields. In polymer networks wherein chain ends crosslink into a network, entanglements become topologically locked and cannot disentangle; however, the unbound ends in our two-chain Kremer-Grest simulations can unwrap. We found that setting $\phi=0.1$ and $N>20$ produced entanglements with $|\varTheta_\alpha|>1$ in $33\%$ of cases, which persisted throughout sampling. The upper limit of $N$ was selected because modeling chains with $N>100$ incurs prohibitive wall-clock times that scale as $\mathcal{O}(N^3)$. This is because -- although wall-clock time scales roughly as $\mathcal{O}(N)$ in LAMMPS -- required sampling times scale as $\mathcal{O}(N^2)$ due to increases in the Rouse relaxation time ($\tau_R \propto N^2 \tau_0$ \cite{stukalin_self-healing_2013}) needed to sample chains' conformational states. This not only sets the upper limit of $N$ investigated using CGMD, but also motivates the use of computationally reduced DNMs.  

Chain stretches were prescribed by randomly selecting a targeted, normalized chain end-to-end distance, $r^* = r/Nb$, from the uniform distribution in the range $r^* \in [0.1,0.9]$. Note that we use $r^*=r/Nb$ in lieu of the commonly applied chain stretch, $\lambda_c = r/\sqrt N b$ \cite{wagner_mesoscale_2022, wagner_network_2021}, because it enables direct comparison of results for different values of $N$. We found that when $r^*<0.1$, $\sim 95\%$ of chains disentangled (i.e., $|\varTheta|<1.0$) before the Rouse time could be reached. Therefore, we maintained $r^*\geq 0.1$. We also found that for $r^*>0.9$, chains undergo enthalpic bond stretching \cite{mulderrig_statistical_2023}, which is inconsistent with Kuhn and Gru\"n's assumption of finitely extensible Kuhn segments used later in our DNMs.

Stretch was introduced by applying the affine, isotropic deformation gradient $\bm F = \lambda_I \bm I$ of \textbf{Section~\ref{sec:ModelingEntangledChains}} where $\lambda_I = \min (N_i b \cdot r^*/r_i )$ is the minimum global stretch required to achieve $r^*$ in either chain. Here, $N_i$ and $r_i$ are the number of Kuhn segments and initial end-to-end length of chain $i$, respectively, where $i\in \{1,2\}$. To maintain numerical stability, stretch was increased linearly in time following $\lambda(t) = 1 + (\lambda_I - 1)(t/t_f)$ for $0 \le t \le t_f$, with $t_f = 10^{-3} \tau_0$ representing the time at which stretch $\lambda_I$ is reached. During stretching, the ends of each chain were held stationary while their intermediate beads were relaxed according to Eq.~\ref{eq:position}. Note that this approach drives one chain to the target stretch, $r^*$, while the other chain achieves a stretch below $r^*$. The stretch from the latter chain was retroactively measured so that its data could be added to the data set. 

Once the chains were stretched, loading history was erased by fixing their free ends as if crosslinked into a greater network and relaxing their intermediate beads following Eq.~\eqref{eq:position}. Relaxation was conducted for a duration of $3\bar \tau_R$, where $\bar \tau_R=\bar N^2\tau_0$ is the Rouse time \cite{stukalin_self-healing_2013} of a chain containing the average number of Kuhn segments, $\bar N=60$. 

    % fig 7

To evaluate whether entropic forces pass through the COE positions of Eq. \eqref{eq:COE}, we sampled the instantaneous unbalanced forces acting on each fixed end point of the two chains with a sampling rate of $\tau_0^{-1}$, as illustrated in Fig.~\ref{fig:Primitive_Path_Alignment}(a). These represent the forces that would be transmitted to a polymer network if the chains were crosslinked at their ends. Sample chain configurations and corresponding force fields projected onto a 2D plane are provided for reference in \textbf{SI Section S2} (Fig.~S3). We then projected the direction of forces, $\hat {\bm f}$, onto the direction of corresponding primitive path vectors, $\hat{\bm r}_p$, and averaged the results over time (denoted by $\langle \bm{\hat f}\cdot \bm{\hat r}_p \rangle$). If $\langle \bm{\hat f} \cdot \bm{\hat r}_p \rangle =1$, then time-averaged chain forces align directly with COE positions. Results are displayed in Fig.~\ref{fig:Primitive_Path_Alignment}(b) for averaging times of $\{ 0.03,0.3,3 \}\bar \tau_R$. 

Fig.~\ref{fig:Primitive_Path_Alignment}(b) reveals that for timescales well below $\bar \tau_R$, time-averaged chain forces align with COE positions only at high stretches, which is expected since highly stretched chains exhibit fewer conformational degrees of freedom and thus shorter relaxation times are required to sample their ergodic states. However, as relaxation time increases, dependence on chain stretch steadily diminishes and for hold times above $\bar \tau_R$ entropic forces nicely align with the postulated COE positions of Eq. \eqref{eq:COE} regardless of chain stretch. 

To further assess whether entanglements act as effective load-bearing junctions, we compare the force-extension relations ($\bar f$ versus $r^*_p = r_p / N_p b$) from the Kremer-Grest simulations to the statistical mechanics solution, $f=-\partial \psi/\partial r$ from Eq.~\eqref{eq:Pade} in  Fig.~\ref{fig:Primitive_Path_Alignment}(c). These models show strong agreement ($R^2 = 0.99$) for $r_p^* < 0.9$, indicating that the segment between a fixed end and the COE effectively behaves as a freely jointed chain of finitely extensible Kuhn segments. This supports our hypothesis that COE acts as the load-transmission position at which the entangled chains are effectively partitioned into primitive paths of lengths, $N_p b$.

%%%%%%%%%%%%%%%%%%%%%%%%%%%%%%%%%%%%%%%%%%%%%%%%%%%%%%%%%%%%%%%%%%%%%%%%%%%%%%%%%%%%%%%%%%%%%%%%%%%%
\section{\label{sec:COEDiffusion}COES EXHIBIT ANISOTROPIC, STRETCH-DEPENDENT MOBILITY}

\begin{figure}[!htbp]
        \includegraphics[width=\columnwidth]{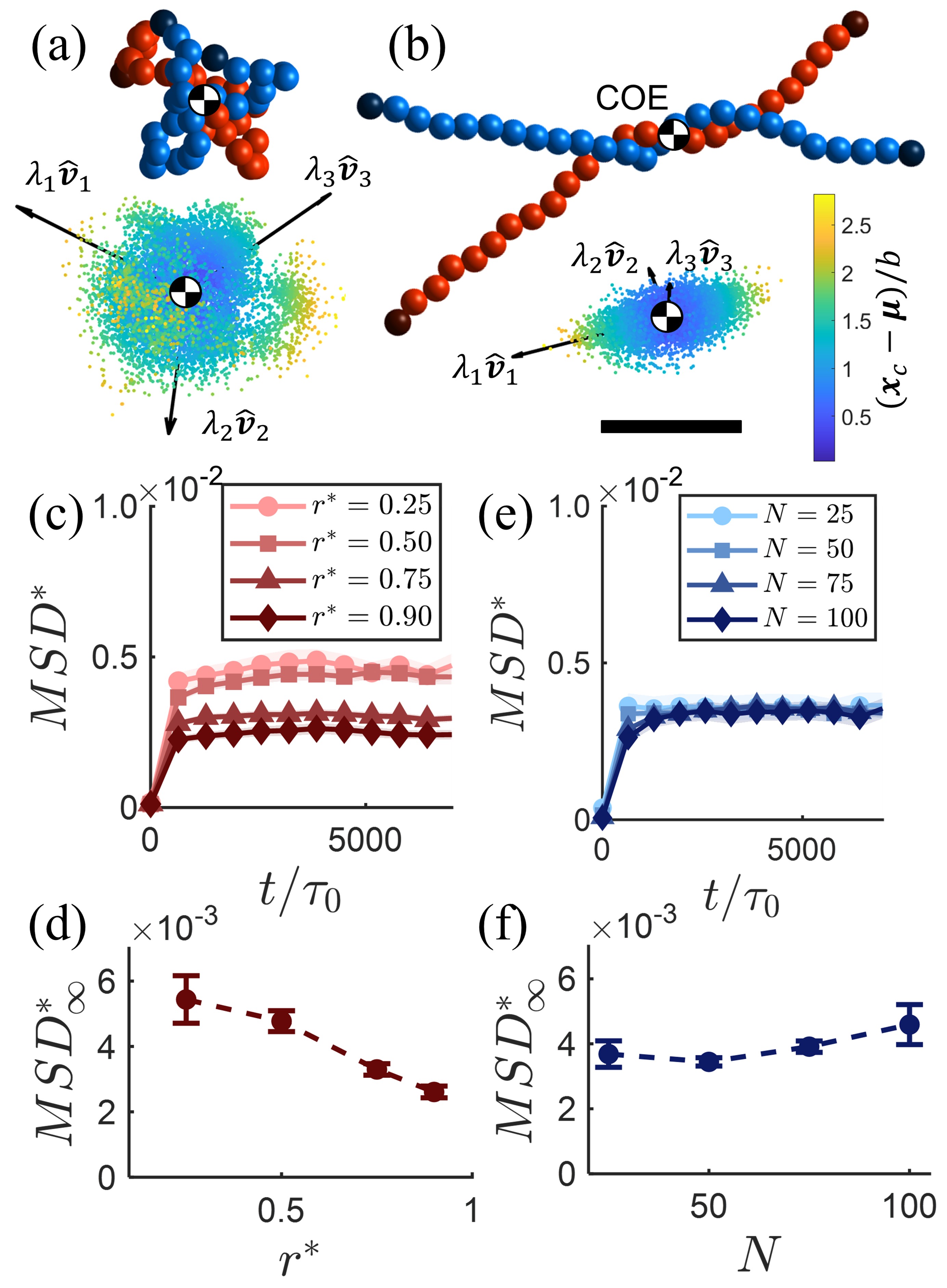}
        \caption{\label{fig:COE_diffusion}
        \textbf{Diffusive Behavior of Entanglements.} 
        \textbf{(a)} $r^* = 0.1$ and \textbf{(b)} $r^* = 0.9$ (top). 
        The corresponding distributions of all sampled COE positions are shown below each snapshot. The mean COE position $\bm{\mu}$ is marked by the black and white symbol, and the eigenvectors of the positional covariance are indicated by arrows. 
        Colors in the point clouds indicate COE distance from the mean position, $\|\bm{x}_{c,i} - \bm{\mu}\|/b$.
        The scale bar represents $5b$. 
        \textbf{(c,d)} $MSD^*$ of the COE, with respect to time, $t/\tau_0$, for \textbf{(c)} various stretches $r^*$ when $N \in [20,100]$ and \textbf{(d)} various chain lengths $N$ when $r^* \in [0.1,0.9]$. 
        \textbf{(d,f)} Corresponding plateau values $MSD^*_\infty$,versus \textbf{(d)} stretch $r^*$ and \textbf{(f)} chain length $N$.
        Error bars in \textbf{(c-f)} denote standard error of the mean.}
    \end{figure}

The alignment and force--stretch results in \textbf{Section~\ref{sec:ReliabilityOfPrimitivePaths}} validate primitive path assumptions at timescales exceeding the Rouse time, and also verify that our representation of COEs captures the effective location of inter-chain load transmission due to entanglement. However, for the purposes of distilling CGMD networks into equivalent DNM representations, we also aim to understand how COE positions fluctuate in space and time. Figs.~\ref{fig:COE_diffusion}(a-b) illustrate two examples of stochastically generated chains at stretches of $r^*=0.1$ and $r^*=0.9$. Their 3D distributions of COE positions are presented as point clouds directly beneath them. To spatially characterize these distributions, we compute their covariance tensors as $\bm C^\alpha_{c}=\frac{1}{n-1}\sum_k^n \bm{q}_k \bm{q}_k^T$, where $\bm{q}_k = \bm{x}_{c,k}^\alpha - \bm{\mu}_{c}^\alpha$  and $\bm{\mu}_{c}^\alpha$ is the mean COE position of entanglement $\alpha$ over all $k\in [1,n]$ observations. The eigenvectors of $\bm C_c^\alpha$ indicate the principal directions of entanglement mobility, while their magnitudes represent the positional variance in these directions. Furthermore, the ratios between the eigenvalues of $\bm C_c^\alpha$ characterize anisotropy in entanglement mobility (with ratios of unity indicating isotropy). 

Figs.~\ref{fig:COE_diffusion}(a-b) display the eigenvectors of $\bm C_c^\alpha$ for the two-chain systems  (black arrows). These illustrate how entangled chains at higher stretches tend to exhibit lower variance, but higher anisotropy in COE positions. Specifically, the maximum eigenvectors of $\bm C_c^\alpha$ tend to align with the orientation of the double helical entanglement structure (e.g., Fig.~\ref{fig:COE_diffusion}(b)). Quantitatively, this diffusional anisotropy is confirmed by a small increase in maximum-to-minimum eigenvalue ratios of $\bm C_c^\alpha$ as chain stretch is increased (see \textbf{SI Section S3}, Fig.~S4). This provides evidence that for highly stretched chains (which maintain relatively sustained inter-chain contact) entanglement mobility is predominantly governed by reptative chain sliding along contour paths. In contrast, entanglement mobility of chains at low stretch is governed by relatively unconstrained chain diffusion kinetics. 

    % fig 8
    
Regardless of its origins, anisotropic COE mobility does not significantly alter the distillation procedure and DNM mechanical responses targeted here for a few key reasons. First, even the largest principal fluctuation lengths are small (on the order of $\sim 2b$) compared to the contour lengths ($Nb$) of the chains investigated, so that realignment of primitive paths due to COE fluctuations is also small. Second, anisotropy in positional COE distributions does not affect the time-averaged entropic forces of polymer chains, which pass through the average COE positions as demonstrated in \textbf{Section~\ref{sec:ReliabilityOfPrimitivePaths}}. Third, any effect anisotropic COE mobility has on the primitive path vector population — and thus on stress — cancels out in large, isotropic polymer networks where COE orientations are randomly and uniformly distributed. Therefore, while we may incorporate entanglement orientations as vertex state variables in future DNM efforts, here we omit this trait for parsimony.

To characterize the spatiotemporal exploration of entanglements, we also computed
the normalized mean-squared displacement (MSD) of the COE position, $MSD^* = \langle \Delta x_c^{2}(t)\rangle/(Nb)^2$. Here, normalization by $(Nb)^2$ provides the areal scale of COE oscillation relative to the overall size of the chains. Results are shown in Fig.~\ref{fig:COE_diffusion}(c-f). Fig.~\ref{fig:COE_diffusion}(c) depicts the ensemble-averaged $MSD^*$ versus time for various chain stretches of lengths $N \in [20,100]$, while Fig.~\ref{fig:COE_diffusion}(d) depicts the plateau value, $MSD^*_\infty$, versus chain stretch. As $r^*$ increases, $MSD^*_\infty$ decreases monotonically. Indeed, at $r^*=0.9$, $MSD^*_\infty = (2.6\pm 0.2) \times 10^{-3}$, which is only half of the value at $r^*=0.1$, $MSD^*_\infty = (5.4 \pm 0.7) \times 10^{-3}$, indicating that as chains are taut, their entanglement locations become more clearly defined as expected. 

Fig.~\ref{fig:COE_diffusion}(e) depicts $MSD^*$ versus time for various two-chain systems with various values of average chain length, $N$, for chains stretched to $r^* \in [0.1, 0.9]$, while Fig.~\ref{fig:COE_diffusion}(f) shows $MSD^*_\infty$ versus $N$. Notably, $MSD$ collapses when normalized (Fig.~\ref{fig:COE_diffusion}(e)) with an average value of $MSD^*\sim 3.9\times 10^{-3}$ and no significant variation in $MSD^*_\infty$ with respect to $N$ (Fig.~\ref{fig:COE_diffusion}(f)). This suggests that it is the available contour length on either side of an entanglement that enables its positional diffusion and provides further evidence that COE movement is driven by bead motion along the contour paths of the chains.

%%%%%%%%%%%%%%%%%%%%%%%%%%%%%%%%%%%%%%%%%%%%%%%%%%%%%%%%%%%%%%%%%%%%%%%%%%%%%%%%%%%%%%%%%%%%%%%%%%%%
\section{\label{sec:DNM_Mechanical_Response} DISTILLED DNMS REPRODUCE CGMD STRESSES AND NETWORK STRUCTURES}%:\protect\\

The results of \textbf{Sections \ref{sec:ReliabilityOfPrimitivePaths}} demonstrate that entropic forces pass through the COE positions defined in Eq.~\eqref{eq:COE} at timescales above the Rouse relaxation time (i.e., $t>\tau_R$) at all values of chain stretch and length. Importantly, this is not restrictive for intended DNM applications, which inherently target longer timescales where such conditions are naturally satisfied. Furthermore, the results of \textbf{Section~\ref{sec:COEDiffusion}} show that COEs oscillate around relatively small regions within $7\%$ of chain's overall contour length. These results physically motivate the distillation procedure of \textbf{Section~\ref{sec:DistillingPolymerNetworks}}. Here we quantitatively assess whether this procedure generates DNMs that accurately represent the topologies and mechanical responses of their parent CGMD networks. 

\begin{figure}[!htbp]
        \includegraphics[width=\columnwidth]{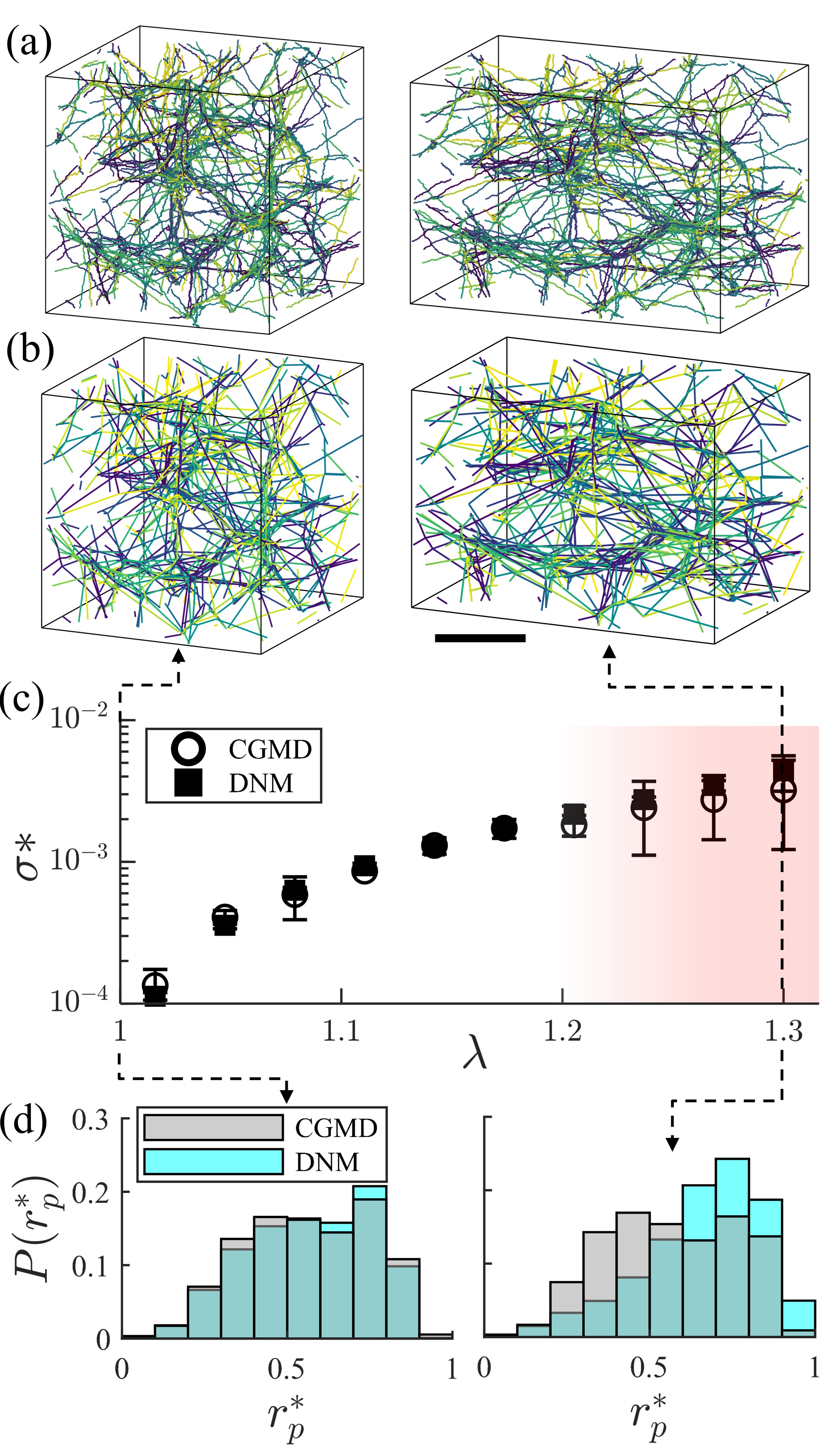}
        \caption{\label{fig:virial_stress}
        \textbf{Characterizing DNM distillation accuracy.}
        \textbf{(a)} CGMD and \textbf{(b)} DNM RVEs at $\lambda = 1.0$ (left) and $\lambda = 1.3$ (right). Different colors represent distinct molecular chains. Scale bar represents $100b$. 
        \textbf{(c)} Ensemble-averaged stress, $\sigma^*$, with respect to stretch, $\lambda$, for $n=5$ randomly generated network configurations comprised of $\mathcal{N}=50$ chains with $N=1000$. Circles and squares indicate data from the CGMD model and DNMs, respectively ($R^2=0.98$). Error bars denote standard error of the mean. The red shaded region indicates the region where CGMD and DNM responses begin to deviate due to CGMD chain sliding.        \textbf{(d)} PDFs of primitive-path stretch, $r_p^*$, for CGMD (grey) and DNM (cyan) networks at $\lambda=1.0$ (left) and $\lambda = 1.3$ (right).}
    \end{figure}

First, we initiated CGMD networks composed of $\mathcal{N} = 50$ Kremer-Grest chains with $N = 1,000$ Kuhn segments at a packing fraction of $\phi = 0.5$ following the procedure of \textbf{Section~\ref{sec:ModelingEntangledChains}}. This generated periodic RVEs with on the order of $1,400$ entanglements. Once generated, the RVEs were stretched via the affinely applied, isotropic deformation gradient, $\bm F = \lambda_I \bm I$ (with $\lambda_I = 5$), and then relaxed for $t=10^3\tau_0$ (Fig.~\ref{fig:virial_stress}(a)). For the purposes of validating the distillation procedure -- as opposed to the DNMs, themselves (which have been extensively explored in other works \cite{gusev_molecular_2024,bernhard_phantom_2025,assadi_nonaffine_2025,huang_topological_2025}) -- we simply fixed each chain's ends so that they followed applied deformations affinely. Only the intermediate bead positions were updated according to Eq.~\eqref{eq:position} during relaxation and deformation. This approach adequately mimics permanent elastomers and gels without requiring the implementation of complex crosslinking algorithms. %and advanced DNM mechanisms such as the reptative sliding developed by Assadi et al. (2025) \cite{assadi_nonaffine_2025}. 

Once the CGMD networks were initiated, we constructed corresponding DNMs using the topological distillation framework of \textbf{Section~\ref{sec:DistillingPolymerNetworks}} (Fig.~\ref{fig:virial_stress}(b)). Using LAMMPS, we then subjected both the original CGMD networks and their corresponding DNMs to incompressible uniaxial tension following the deformation gradient, $\bm{F}=\mathrm{diag}\left(\lambda,\lambda^{-1/2},\lambda^{-1/2}\right)$, where $\lambda=1 + \dot\epsilon t$ and $\dot \epsilon$ is the nominal strain rate. Since our aim is simply to evaluate the physical distillation accuracy, the DNM implemented here neglects entanglement sliding \cite{assadi_nonaffine_2025,huang_topological_2025} and viscous drag calibration \cite{wagner_foundational_2025}. Therefore,  deformation was carried out up to relatively small stretches of $\lambda = 1.3$, and the nominal strain rate was set to $1.5\times 10^{-4} \tau_0^{-1}$ to achieve quasi-static loading conditions (Figs.~\ref{fig:virial_stress}(a-b)).

    %fig 9

To evaluate mechanical response predictions, we computed the potential term of the virial stress according to:
 
     \begin{equation}
        \bm{\sigma}=\frac{1}{2V}\sum_{k,l}\left(\bm{r}_{kl}\otimes\bm{f}_{kl}\right),
        \label{eq:virial_stress}
    \end{equation}
    
\noindent where $V$ is the system volume, and $\bm r_{kl}$ and $\bm f_{kl}$ represent the set of pairwise end-to-end vectors and interaction forces of all Kuhn segments or primitive paths in the Kremer-Grest model and DNM, respectively. The ensemble average of normalized axial stress $\sigma^*=\sigma_{11} b^3/k_B T$ (where $\sigma_{11}$ is the normal stress in the loading direction) is plotted with respect to $\lambda$ in Fig.~\ref{fig:virial_stress}(c) for $n=5$ network samples. 

The distilled DNMs predict virial stress response in good agreement with that of the CGMD over the stretch range considered here ($R^2=0.98$), but with significantly reduced computational cost. The DNMs contain $\sim$97\% fewer degrees of freedom than their parent CGMD networks, reducing the model from $50{,}000$ polymer beads to approximately $1{,}400$ entanglement vertices. This results in a $\sim$99\% reduction in CPU hours ($512$ CPU hours for the CGMD models versus $4.5$ for the DNMs) when both models were run in LAMMPS on sixteen Intel\textsuperscript{\textregistered} Xeon\textsuperscript{\textregistered} Gold 6534 CPU cores using message passing interface. Additionally, the DNM outputs files with a $\sim$97\% reduction in storage requirements as compared to the CGMD model ($950$ MB and $25$ MB per simulation for the CGMD and DNM, respectively).

While small-strain mechanical predictions are in particularly good agreement, the DNM response becomes slightly stiffer than that of the CGMD model at high stretches (red shaded region in Fig.~\ref{fig:virial_stress}(c)). To elucidate the micromechanical origins of this deviation, we compared the probability distribution functions (PDFs) of primitive-path stretch, $r_p^*$, from the CGMD networks and DNMs at $\lambda=1.0$ and $\lambda=1.3$ (Fig.~\ref{fig:virial_stress}(d)). At $\lambda=1.0$, the primitive path stretches are in close agreement between models, indicating good agreement at the time of construction. Minor deviations are observed due to thermal fluctuations in the CGMD model. After stretching to $\lambda=1.3$ the DNM distribution shifts toward larger values of $r_p^*$, whereas the CGMD distribution remains relatively unchanged. This shift to higher primitive path stretches partly explains the slightly stiffer DNM response observed in Fig.~\ref{fig:virial_stress}(c) at high stretches and arises due to lack of chain sliding (i.e., redistribution of Kuhn segments across entanglements). Another reason for the slightly higher DNM stress response at high stretch is a lack of enthalpic bond stretching and, as a result, chain softening \cite{mulderrig_statistical_2023, zhu_stretching_2025} through Eq.~\eqref{eq:Pade}. This is evidenced by the slightly higher fractions of CGMD chains stretched to $r_p^*>0.9$ in Fig.~\ref{fig:virial_stress}(d) even in the unstretched networks. While neither of these effects have been incorporated into the DNM here, they have been \cite{assadi_nonaffine_2025,huang_topological_2025} or can be \cite{mulderrig_statistical_2023} readily included in such models.

%%%%%%%%%%%%%%%%%%%%%%%%%%%%%%%%%%%%%%%%%%%%%%%%%%%%%%%%%%%%%%%%%%%%%%%%%%%%%%%%%%%%%%%%%%%%%%%%%%%%
\section{\label{sec:Conclusion} SUMMARY AND CONCLUSION}
We here developed a topological method of identifying mechanically relevant entanglements along the backbones of discretely modeled polymer chains. To contend with the ambiguity and transience of entanglements, we introduced a topological framework that not only defines entanglement degree through a local version of the linking number, but also defines a geometric COE position at which entanglements effectively transmit entropic forces. Employing Kremer-Grest models, we validated our hypothesis that these COE positions act as the spatial coordinates through which time-averaged primitive path forces pass. Furthermore, we confirmed that the force-extension relations of these primitive paths are well represented  by the Pad\'{e} approximation \cite{cohen_pade_1991} of Helmholtz free energy for freely jointed chains as introduced by Kuhn and Gr\"{u}n \cite{kuhn_beziehungen_1942}.

These results hold across all investigated chain stretches and lengths so long as hold times exceeded the Rouse time. This confirms that primitive path analysis should be restricted to models operating at relatively large timescales (exceeding chain relaxation times). DNMs, which are specifically employed towards bridging scales between molecular physics and macroscale mechanical properties, fit this criterion. As such, we leveraged our findings to introduce a physically motivated distillation procedure that converts entangled CGMD networks into equivalent DNMs. We found that the DNMs predict virial stress response in good agreement with CGMD predictions up to 30\% engineering strain even without entanglement sliding as seen in Assadi et al. (2025) or Huang et al. (2025). They achieve this accuracy while reducing computational cost from CGMD by $\geq97\%$. Moving forward, our findings and distillation procedures may be used as part of physics-based, multiscale modeling pipelines to map structure-property relations of entangled polymers and facilitate predictive material design of polymers, metamaterials, and other entangled network structures. \newline 
~

%%%%%%%%%%%%%%%%%%%%%%%%%%%%%%%%%%%%%%%%%%%%%%%%%%%%%%%%%%%%%%%%%%%%%%%%%%%%%%%%%%%%%%%%%%%%%%%%%%%%
\section*{Acknowledgments}
This work was supported by the State University of New York (SUNY) at Binghamton University and the United States National Science Foundation (NSF) under Grant No. CAREER-2539455. This content is solely the responsibility of the authors and does not necessarily represent the official views of SUNY, Binghamton University, or NSF. This work used the high-performance (Spiedie) computing cluster at Binghamton University to produce results. The authors would like to thank the Watson Computing Staff at Binghamton University for their oversight and management of these computational resources.

%%%%%%%%%%%%%%%%%%%%%%%%%%%%%%%%%%%%%%%%%%%%%%%%%%%%%%%%%%%%%%%%%%%%%%%%%%%%%%%%%%%%%%%%%%%%%%%%%%%%
\section*{Data Availability}
The source codes used to generate the results of this manuscript are curated and available at \url{https://github.com/sifat-mottaqin/primitive-path-analysis-polymer-networks} \cite{mottaqin_pp_code_2026}. Data sets are available from the authors upon reasonable request.
%%%REFERENCES%%%
\bibliography{citations} %You need to replace "rsc" on this line with the name of your .bib file
\bibliographystyle{elsarticle-num-names} %the RSC's .bst file

\end{document}